\documentclass[twocolumn, prl]{revtex4}
\pdfoutput=1
\usepackage{graphicx}
\usepackage{epsfig}
\begin{document}
\title{Fate of a Bose-Einstein Condensate in the Presence of Spin-Orbit Coupling}
\author{ Qi Zhou$^{1}$ and Xiaoling Cui$^{2}$}
\affiliation{$^{1}$Department of Physics, The Chinese University of Hong Kong,  Shatin, New Territories, Hong Kong\\
$^{2}$ Institute for Advanced Study, Tsinghua University, Beijing 100084, China}

\date{\today}

\begin{abstract}
Intensive theoretical studies have recently predicted that a Bose-Einstein condensate will exhibit a variety of novel properties if spin-orbit coupling is present. However, an unambiguous fact has also been pointed out: Rashba coupling destroys a condensate of noninteracting bosons even in high dimensions. Therefore, a conceptually important question arises as to whether or not a condensate exists in the presence of interaction and a general type of spin-orbit coupling. Here we show that interaction qualitatively changes the ground state of bosons under Rashba spin-orbit coupling. Any infinitesimal repulsion forces bosons either to condense at one or two momentum states or to form a superfragmented state that is a superposition of infinite numbers of fragmented condensates. The superfragmented state is unstable against the anisotropy of spin-orbit coupling in systems with large numbers of particles, leading to the revival of a condensate in current experiments.
\end{abstract}

\maketitle

Spin-orbit coupling(SOC) is the underlying mechanism for many
fundamental quantum phenomena, ranging from
the atomic fine structure to the newly discovered novel properties   of
 topological
insulators\cite{TIS,TI}.The recent realization of synthetic SOC for
neutral alkali atoms in laboratories  provides physicists a new
platform to study the effects of SOC in many-body systems, in which
a wide range of parameters can be well controlled in
experiments\cite{Ian, Martin, Chen, Zhang}. In particular, spin-orbit
coupled bosons offer physicists a unique opportunity to explore how
SOC may manifest itself at the macroscopic level. As it is
known as a textbook result that bosons naturally form a condensate
at the ground state in three and two dimensions, intensive
theoretical efforts have predicted a number of
macroscopic quantum phenomena exhibited
by a  Bose-Einstein condensate in the presence of
SOC\cite{Boson, Zhai, Galitski, Wuc, Barnett, Hu, Stringari, Santo,
Ozawa, Ueda}.

On the other hand, a significant effect of SOC on
non-interacting bosons has also been realized recently. It was pointed
out that some types of SOC may completely destroy a non-interacting
condensate at zero or any finite temperatures even in high
dimensions \cite{LXJ, Ozawa2,
Depletion}. To see this effect, one can start from the
single-particle Hamiltonian for spin-orbit coupled bosons, which can
be written as
\begin{equation}
\mathcal{K}=\sum_\sigma \int d^D{r} \hat{\Psi}_\sigma^\dagger({\bf r})\left(-\frac{\hbar^2{ \nabla}^2}{2M}\right)\hat{\Psi}_\sigma({\bf r})+H_{SOC},
\end{equation}
where $D$ is the dimension, $\sigma=\uparrow, \downarrow$, $\hat{\Psi}^\dagger_\sigma({\bf r})$($\hat{\Psi}_\sigma({\bf r})$)  is the creation (annihilation) operator at ${\bf r}$ and $M$ is the mass. The part of SOC in the Hamiltonian is described by
\begin{equation}
H_{SOC}=-i\int d^D{r}\hat{\Psi}^\dagger_\uparrow({\bf r})(\lambda_x  \partial_x-i  \lambda_y \partial_y) \Psi_\downarrow({\bf r})+c.c.,
\end{equation}
where $\lambda_{x,y}$ is the coupling strength.  For $D=3$ the
low energy part of single-particle spectrum can be written as
\begin{equation}
\epsilon_{\bf k}=\frac{\hbar^2}{2M}(|{\bf
k}_{\bot}|^2+k_z^2)-\sqrt{\lambda_x^2k_x^2+\lambda_y
k_y^2},\label{SE}
\end{equation}
where ${\bf k}=(k_x, k_y, k_z)$ is the momentum and ${\bf
k}_\bot=(k_x, k_y)$. For $D=2$, one can simply set $k_z=0$ in
Eq.(\ref{SE}). Among all the configurations of SOC, Rashba coupling
corresponding to $\lambda_x=\lambda_y=\lambda$ is of particular
interest. In both $D=2$ and $D=3$,  the kinetic energy minimum  under Rashba coupling
 becomes a circle in the x-y plane
with radius $|{\bf k}_\bot|=k_0\equiv M\lambda/\hbar^2$, which means an infinitely degeneracy of the single-particle ground state.
Correspondingly, the low-energy Density of States  becomes that in
$(D-1)$ dimension in the absence of SOC.  Therefore, a non-interacting
condensate is completely destroyed at zero temperature for $D=2$\cite{ Depletion} and at any
finite temperature for $D=3$\cite{LXJ, Ozawa2}.

Whereas the disappearance of a non-interacting condensate under
Rashba coupling is rather clear,  a few fundamental questions remain unanswered so far. (i) What is the ground state of interacting bosons,  a trivial uncondensed state, an ordinary condensate or an exotic many-body state  with novel correlations? (ii) If it is a condensate  or an exotic state, how does the uncondensed state of non-interacting bosons change to these states when interaction is turned on? (iii) What is the effect of anisotropy in spin-orbit coupling? In this Letter, we shall present answers to all above questions (i-iii). 

We first note that the infinite degeneracy of
single-particle ground states makes interaction effects highly
non-perturbative. To determine the many-body ground state amounts to
selecting the state with the lowest interaction energy from an
infinitely degenerate subspace, which is known to be a challenging
problem, for instance, two-dimensional electrons in Quantum Hall regions\cite{QH}.  For studies on  spin-orbit coupled bosons in the literature, a mean field approach has been adopted
to compare energies of a special class of states that can be described
by condensate wave functions\cite{ Zhai, Galitski, Wuc, Barnett, Hu,
Santo, Ozawa}. However, with the observation that SOC may completely
destroy a condensate, one naturally concerns the validity of mean
field approach. In particular, it is unclear whether interaction will lead to exotic many-body ground state other than an ordinary condensate.

 \begin{figure*}[tbp]
\begin{center}
\includegraphics[width=6.3in]{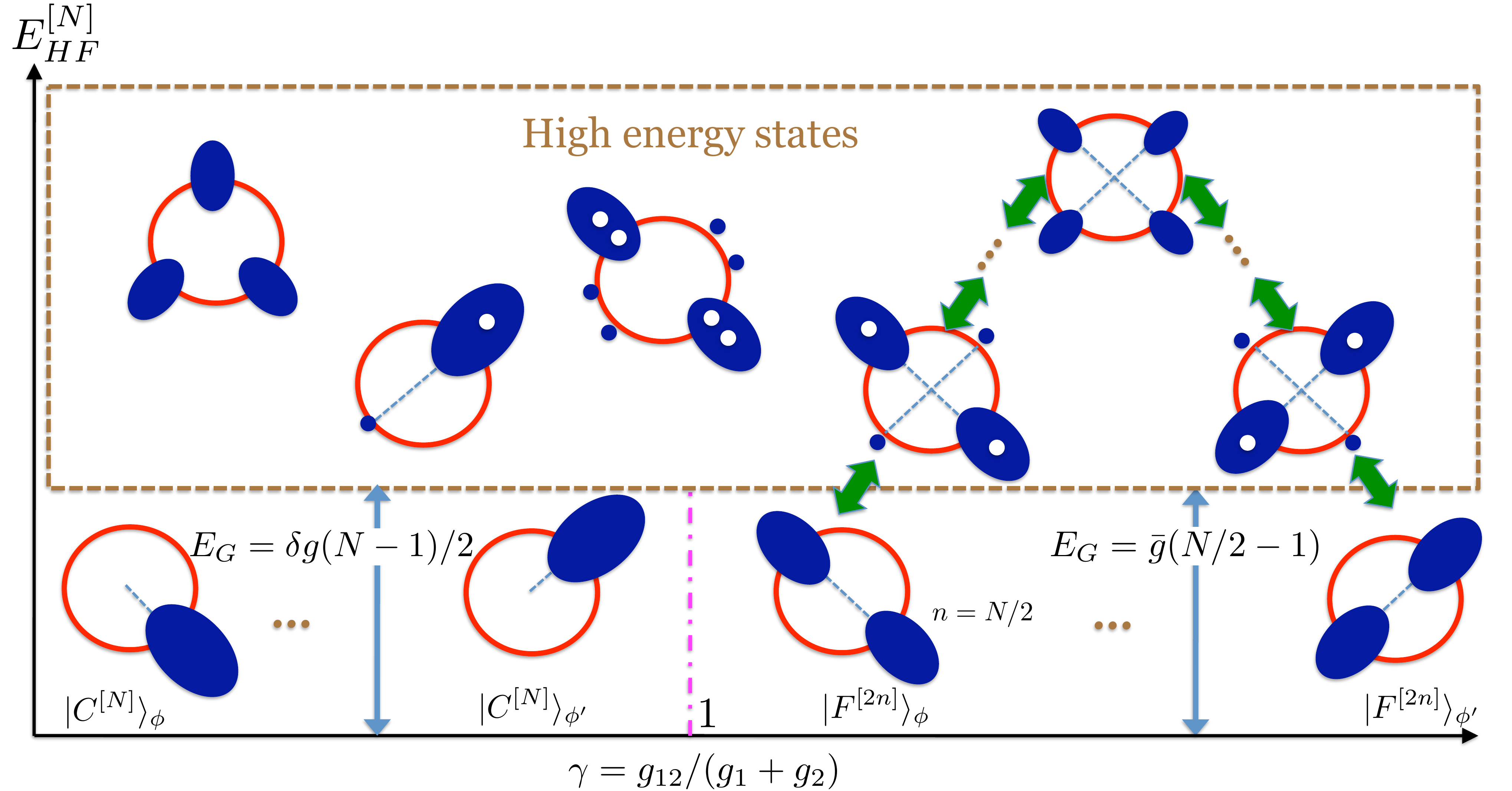}
\end{center}
\caption{Hartree-Fock energy $E_{HF}^{[N]}$ of $N$ particles. Red circle represents the kinetic energy minimum, and blue ellipses represent a large number of bosons occupying the same momentum state. Blue and white dot represent a single boson and a hole in the condensate respectively. Single-particle and fragmented condensate minimize $E_{HF}^{[N]}$ for $\gamma<1$ and $\gamma>1$. They are separated from high energies states by an energy gap $E_G=\bar{g}+N\delta g/2$ or $E_G=\bar{g}({N}/{2}-1)$, where $\bar{g}=(g_1+g_2+g_{12})$ and $\delta g=g_1+g_2-g_{12}$.  Any two fragmented condensates with even particle numbers $N=2n$ are coupled through a sequence of tunneling induced by $n$ steps of scattering, as shown by the green arrows.}
 \end{figure*}

To explore the exact many-body ground state, we expand the interaction between two-component bosons
\begin{equation}
\hat{\mathcal{U}}=\Omega\int d^D{r}(g_1\hat{n}_\uparrow^2({\bf r})+g_2\hat{n}_\downarrow^2({\bf r})+g_{12}\hat{n}_\uparrow({\bf r})\hat{n}_\downarrow({\bf r}))
\end{equation}
 in the basis of single-particle eigenstates, where $\hat{n}_\sigma({\bf r})=\hat{\Psi}^\dagger_\sigma({\bf r})\hat{\Psi}_\sigma({\bf r})$ is the density operator, $\Omega$ is the volume of the system, and $g_{i=1,2}>0 $, $g_{12}>0$ are the intra- and inter-spin repulsion. In this basis, the effective interaction can be formulated as
\begin{equation}
\hat{\mathcal{U}}_L=\sum_{\phi_1,\phi_2,\phi_3,\phi_4} U_{\phi_1,\phi_2}^{ \phi_3,\phi_4}\hat{L}^\dagger_{{\phi_1}}\hat{L}^\dagger_{{\phi_2}}\hat{L}_{{\phi_3}}\hat{L}_{{\phi_4}}/4, \label{EU}
\end{equation}
where
\begin{equation}
U_{\phi_1,\phi_2}^{\phi_3,\phi_4}=g_{12}e^{i({\phi_1}-{{\phi_4}})}+g_{2}+g_{1}e^{i({{\phi_1}}+{{\phi_2}}-{{\phi_3}}-{{\phi_4}})},\label{eU}
\end{equation}
$\hat{L}^\dagger_\phi$ ($\hat{L}_{\phi}$) is the creation (annihilation) operator in the subspace $\mathcal{L}$ composed by all single-particle ground states, and $\phi$ is the polar angle on the circle $|{\bf k}_\bot|=k_0$\cite{Depletion}.

Before presenting the details of our analysis, we first summarize our main results here. (I) any infinitesimal repulsion forces bosons to form a single-particle condensate or a fragmented condensate, distinct from non-interacting systems where a condensate is absent. (II) interaction inevitably builds up a superposition of fragmented condensates with even particle number, and the exact ground state becomes a super-fragmented state beyond the prediction of mean field theory. (III) super-fragmented states collapse in systems with large numbers of particle if a finite anisotropy exists in SOC, leading to the revival of a condensate in current experiments. (I-III)  answer the questions (i-iii) raised before.

We start from the two-body problem that can be solved exactly to illuminate the underlying physics. The Fock states can be written as either $|\phi,\theta\rangle=\hat{L}^\dagger_\phi\hat{L}^\dagger_{\phi+\theta}|0\rangle$ for $\theta\neq0$ or $|\phi,0\rangle=\hat{L}^\dagger_\phi\hat{L}^\dagger_{\phi}|0\rangle/{\sqrt{2}}$. $\theta$ is the relative angle between the two bosons on the circle. The diagonal term of the interaction matrix element, i.e., the Hartree-Fock energy,  $E^{[2]}_{HF}= \langle \phi,\theta|\hat{\mathcal{U}}_L|\phi,\theta\rangle$, where the superscript $[N]$ denotes a $N$-body system, can be written as
\begin{eqnarray}
E^{[2]}_{HF}=\Bigg\{
\begin{array}{ll}
g_{12}(1+\cos\theta)/2+g_{1}+g_{2},& \theta\neq 0\\
(g_{12}+g_1+g_2)/2.&\theta =0\\
\end{array}\label{Eth}
\end{eqnarray}
Eq.(\ref{Eth}) shows that, $|\phi,0\rangle$ and $|\phi,\pi\rangle$ minimize $E^{[2]}_{HF}$ for $\gamma<1$ and $\gamma>1$ respectively, where $\gamma=g_{12}/(g_1+g_2)$. States with $\theta\neq 0, \pi$ always have higher energies. For $\gamma>1$, two bosons occupy two opposite points on the circle, i.e.,  fragmentation occurs. Without SOC, it is well-known that fragmentation is energetically unfavorable due to the cost of Fock energy when bosons occupy different momentum states\cite{Nozieres}. The presence of Rashba coupling suppresses the term in Fock energy that is proportional to $g_{12}$ by a factor of $\cos\theta$. Therefore,  fragmentation may occur at large enough $g_{12}$.

We now consider the off-diagonal term of interaction matrix elements. Note that for $\theta\neq \pi$, each Fock state is characterized by a unique finite  momentum and cannot be coupled by interaction which conserves the total momentum. Therefore, they are eigenstates in the manifold of kinetic energy minimum.  However, all states with $\theta=\pi$ carry zero total momentum and can be coupled by interaction. The Hamiltonian within the zero momentum subspace can be written as
\begin{equation}
H^{[2]}={E}_{\pi}\int_0^{\pi} d\phi \hat{P}^\dagger_{\phi}\hat{P}_\phi+\int_{0}^\pi d\phi d\phi'\mathcal{V}^{[2]}_{\phi,\phi'}\hat{P}^\dagger_{\phi'}\hat{P}_\phi(1-\delta_{\phi,\phi'}),\label{H2}
\end{equation}
where ${E}_{\pi}=g_1+g_2$ is the Hartree-Fock energy for $\theta=\pi$, $P^\dagger_{\phi}=\hat{L}^\dagger_\phi\hat{L}^\dagger_{\phi+\pi}$ is the pair creation operator, and $\mathcal{V}^{[2]}_{\phi,\phi'}=(g_2+g_1e^{2i(\phi'-\phi)})$.   
Eq.(\ref{H2}) is equivalent to to a one-dimension ring with an ``infinite-range" tunneling\cite{IH}. The ground states are infinitely degenerate and can be written as
\begin{equation}
|SF_{\nu}^{[2]}\rangle=\frac{1}{\sqrt{\pi}}\int _0^\pi  d\phi e^{i2\nu\phi}\hat{P}^{\dagger }_{\phi}|0\rangle,\label{G2}
\end{equation}
where $\nu$ is an arbitrary integer other than $0$ and $1$. 

The wave function in Eq.(\ref{G2}) describes a small super-fragmented state composed by superposition of small fragmented condensates.  It is straightforward to derive that the corresponding ground state energy is exactly zero, in spite of the repulsive interaction. As the eigen energy of any state with a finite total momentum is positive, we conclude that the ground state for the two-body problem is always a small super-fragmented state, regardless of the ratio between $g_{12}$ and $g_1+g_2$. The underlying physics is that the superposition of zero momentum states on the circle completely cancels the positive Hartree-Fock energy. This is consistent with the result obtained from  a different approach of renormalizing the interaction, which found that the renormalized interaction between two particles with zero momentum on the circle is reduced to zero\cite{Goldbart, Baym}.  

 Above discussions can be directly generalized to a $N$-body system. For systems with even particle number, a single-particle condensate $
|C^{[N]}\rangle_\phi=\hat{L}_\phi^{\dagger N}|0\rangle/{\sqrt{N!}}$ and a fragmented condensate $
|F^{[N]}\rangle_\phi=\hat{L}^{\dagger N/2}_\phi \hat{L}^{\dagger N/2}_{\phi+\pi}|0\rangle/{(N/2)!} $
minimize $E^{[N]}_{HF}$ for $\gamma>1$ and $\gamma<1$ respectively,  where $\phi$ is an arbitrary phase   reflecting the rotation symmetry in the momentum space.
 For odd particle number $N=2n+1$ ($n=1,2...$),  the fragmented condensate is simply given by $
|F^{[2n+1]}\rangle_\phi=\hat{L}^{\dagger n}_\phi \hat{L}^{\dagger n+1}_{\phi+\pi}|0\rangle/{\sqrt{n!(n+1)!}}$. Any other Fock state in $\mathcal{L}$ costs additional Hartree-Fock energy proportional to the total particle number, as shown in Fig.1. Therefore, all bosons condense at one or two momentum states to minimize $E_{HF}^{[N]}$,  distinct from non-interacting systems where bosons can distribute on the circle $|{\bf k_{\bot}}|=k_0$ arbitrarily.
 
Among all above states that minimize Hartree-Fork energy,  $|C^{[N]}\rangle_\phi$ or $|F^{[2n+1]}\rangle_\phi$ with different values of $\phi$ cannot be mixed with each other, due to the constraint of total momentum conservation. However, such a constraint is absent for $|F^{[2n]}\rangle_\phi$, as all of them have zero total momentum and any two of them are inevitably coupled by a series of scattering, as demonstrated in Fig. 1.  This type of tunneling is known to be crucial in the presence of degenerate ground states, and produces macroscopic or mesoscopic quantum coherence in various systems\cite{Leggett,nanom, DL, Pethick}. In our case, it fundamentally changes the ground state structure when fragmentation occurs with even particle number.

We derive an effective Hamiltonian in the subspace composed by all fragmented condensates (See Supplementary material),
\begin{equation}
H^{[2n]}_{eff}=\int_0^{\pi} d\phi \mathcal{E}_{F}^{[2n]}\hat{O}^\dagger_{\phi}\hat{O}_\phi+\int_{0}^\pi d\phi d\phi'\mathcal{V}_{\phi,\phi'}\hat{O}^\dagger_{\phi'}\hat{O}_\phi(1-\delta_{\phi,\phi'}),\label{Heff}
\end{equation}
where $\hat{O}^\dagger_\phi=\hat{L}^{\dagger n}_\phi\hat{L}^{\dagger n}_{\phi+\pi}/{n!}$ is the creation operator for a fragmented condensate, and
\begin{equation}
\mathcal{V}_{\phi,\phi'}=(-1)^{n-1}\frac{(g_2+g_1e^{i2(\phi'-\phi)})^{n}}{(g_1+g_2+g_{12})^{n-1}}n^2. \label{Vn}
\end{equation}
The diagonal term $\mathcal{E}_{F}^{[2n]}=E^{[2n]}_{HF}-\mathcal{E}'$, where $E^{[2n]}_{HF}$ is the Hartree-Fock energy and $\mathcal{E}'$ is the diagonal part of the correction to the energy due to the coupling between $|F^{[2n]}\rangle_\phi$ and high energy states(See supplementary materials). As $\mathcal{E}'$ is independent on $\phi$, it does not affect the structure of ground state wave function.

Similarly to $H^{[2]}$ discussed before,  $H^{[2n]}_{eff}$ is also analytically solvable(See Supplementary Materials). For the simplest case with $g_1=0$,  with an ``infinite-range" constant tunneling  $t=(-1)^{n-1}{{g_2^{n}}n^2}/{(g_2+g_{12})^{n-1}}$\cite{IH}, the eigenstates can be written as
\begin{equation}
|SF_{\nu}^{[2n]}\rangle=\frac{1}{\sqrt{\pi}}\int _0^\pi  d\phi e^{i2\nu\phi}\hat{O}^{\dagger }_{\phi}|0\rangle, \label{S}
\end{equation}
where $\nu$ is an integer, with the corresponding eigenenergy
 \begin{equation}
 E_\nu^{[2n]}=\mathcal{E}_{F}^{[2n]}+(-1)^{n-1}\frac{{g_2^{n}}n^2}{(g_2+g_{12})^{n-1}}(\pi \delta_{\nu, 0}-1).\label{ESC}
 \label{E1}
\end{equation}
As the reduced single-particle density matrix $\langle  \hat{L}^{\dagger }_{\phi}\hat{L}^{}_{\phi'} \rangle$ of $|SF_{\nu}^{[2n]}\rangle$ is zero, and the reduced $2n$-particle density matrix $\langle  \hat{L}^{\dagger n}_{\phi}\hat{L}^{\dagger n}_{\phi+\pi}\hat{L}^{n}_{\phi'}\hat{L}^{n}_{\phi'+\pi} \rangle$ has a unique macroscopic eigenvalue $(n!/\pi)^2$, $|SF_{\nu}^{[2n]}\rangle$ is a super-fragmented state\cite{DL}.  The super-fragmented state we find here has  two exotic features that are absent in previous studies\cite{Leggett,nanom, DL, Pethick}. First, it is composed by an infinite number of macroscopically occupied states other than just a few. Second, each of them is characterized by a winding number $\nu$. In other words, bosons first form a fragmented condensate and then the condensate ``rotates" in the momentum space acquiring a complex phase in the superposition. 
We note that both features originate from the intrigue interplay between Rashba SOC and interaction, which gives rise to an angular-dependent effective interaction as shown in Eq.(\ref{eU}). We have  verified the super-fragmented state as the ground state for a four-body problem using numerical simulations.

It is known that super-fragmented state is fragile in systems with large numbers of particles. In our case, Eq.(\ref{ESC}) shows that the energy difference between the super-fragmented state and the fragmented condensate can be characterized by $\eta=1-{E_\nu^{[2n]}}/{\mathcal{E}_{F}^{[2n]}}$.
For a small  total particle number $N=2n$, $\eta$ takes a large value. For example, one estimates that $\eta\approx 0.61, 0.11, 0.1$ for $N=4, 6,8$, and $g_2/g_{12}=0.9$, which means a large energy gained by forming a super-fragmented state. While this fact  strikingly changes the properties of a few-body system, one notes that  $\eta$ decreases exponentially with increasing particle number $N=2n$, i.e., $\eta\sim\left(1+\frac{g_{12}}{g_2}\right)^{-n+1}$. This is a typical feature of a high order process that requries multiple steps of two-body scattering to couple two degenerate states. When $N\rightarrow \infty$, $\eta$ vanishes, which means a fragmented condensate becomes degenerate with the exact many-body ground state in the thermodynamic limit. In current experiments, $N\sim 10^5-10^6$, one can easily see that $\eta$ is a negligible number.

The tiny $\eta$ for a large value of $N$ indicates that a big super-fragmented state is unstable against external perturbation. For our discussions,  the anisotropy of SOC, which always exists in practice, is naturally such a perturbation. Assuming $\lambda_x>\lambda_y$, Eq.(\ref{SE}) shows that kinetic energy minimums are located at two separate points ${\bf k}= \pm k_{x0}=\pm M\lambda_x/\hbar^2$. We define the circle $|{\bf k}_\bot|=k_{x0}$ as $\mathcal{L}'$, on which the expression for interaction energy is identical to that in Eq.(\ref{Heff}), with $\phi$ replaced by $\varphi=\arg\{\lambda_xk_x, \lambda_yk_y\}$\cite{Depletion}. The kinetic energy in $\mathcal{L}'$ is now a function of $\phi$,
\begin{equation}
K(\phi)=\frac{\hbar^2}{2M}k_{x0}^2-k_{x0}\lambda_x\sqrt{1-\alpha(2-\alpha)\sin^2\phi},\label{Ensp}
 \end{equation}
where $\alpha\equiv (\lambda_x-\lambda_y)/\lambda_x$ characterizes the anisotropy of SOC, and leads to an offset of kinetic energy $\Delta(\alpha)=(K(\pi/2)-K(0))N=\alpha NM\lambda_x^2/\hbar^2$ in $\mathcal{L}'$.

For single-particle condensates and fragmented condensates with odd particle numbers,  any infinitesimal $\alpha$ picks up $|C^{[N]}\rangle_{\phi=0}$ or $|F^{[N]}\rangle_{\phi=0}$ as the ground state, which favors the kinetic energy with no cost of interaction energy.  For super-fragmented states, we define a characteristic anisotropy,
\begin{equation}
\alpha^*=\frac{\hbar^2}{2M\lambda_x^2}\left(\frac{g_2}{g_2+g_{12}}\right)^{n-1}g_2n,\label{astar}
\end{equation}
by setting the kinetic energy offset equal to the strength of interaction induced off-diagonal coupling, i.e., $\Delta(\alpha^*)=t$. For small anisotropy $\alpha\ll \alpha^*$, interaction energy is dominate and super-fragmented state forms. For $\alpha\gg \alpha^*$, to form a superposition of fragmented condensates costs too much kinetic energy, and a single fragmented condensate $|F^{[N]}\rangle_{\phi=0}$ becomes the ground state. A more quantitative analysis is given in Supplementary Materials.  Using Eq.(\ref{astar}), one estimates that $\alpha^*=9\%,4\%, 2\%$ for $N=4, 6,8$, $g_2/g_{12}=0.9$ and $g_2n{\hbar^2}/(2M\lambda_x^2)=0.2$. This means a small super-fragmented state is stable in a finite range of anisotropy. With increasing $N$, $\alpha^*$ decreases exponentially. One can verify  that for $N\sim 10^5-10^6$, the typical particle number in current experiments, $\alpha^*$ is essentially zero. Therefore, one conclude that a super-fragmented state collapses to a fragmented condensate. An additional external potential that breaks the translation invariance shall further change the fragmented condensate to a stripe condensate(See Supplemental Materials).

Whereas we have answered fundamental questions on whether, when and why a condensate exists in the presence of SOC, our work also demonstrates that the coupling between Hartree-Fock energy minimums becomes particularly important in a spin-orbit coupled few-body system. Evidence for this type of coupling has been found in a recent numerical calculation\cite{Hu2}. Provided that significant progresses have been made on manipulating a few atoms in current experiments \cite{Expcl, Greiner} as well as in producing a uniform trap\cite{Zoran}, we hope that our work will stimulate more studies on SOC induced novel ground states of mesoscopic cold atomic systems.

We thank T.L. Ho for helpful discussions. Q.Z. acknowledges start-up support from Department of Physics, The Chinese University of Hong Kong.  X.C. acknowledges support from the Initiative Scientific Research Program of Tsinghua University and NSFC under Grant No. 11104158.

\section{Supplementary material}

In this supplementary material, we present the results on the derivation of Eq.(10) in the main text, the general solution for the effective Hamiltonian $H_{eff}$,  small $N$ systems, effects of anisotropy of spin-orbit coupling, and the evolution from a fragmented condensate to a stripe condensate.

\vspace{0.2in}

{\bf Effective Hamiltonian $H_{eff}^{[2n]}$}

 We define the excited state with $n-m$ pairs of bosons at the angle $\phi$ and $m$ pairs of bosons at $\phi'$ as  $|m\rangle_{\phi,\phi'}$. Within this definition, $|F^{[2n]}\rangle_\phi=|0\rangle_{\phi,\phi'}$,  $|F^{[2n]}\rangle_{\phi'}=|n\rangle_{\phi,\phi'}$. Interaction scatters a pair of bosons out from $|0\rangle_{\phi,\phi'}$ to the momentum states at ${\bf k}'=(k_0,\phi')$ and $-{\bf k}'=(k_0,\phi'+\pi)$. The Schrodinger equation satisfied by $|0\rangle_{\phi,\phi'}$ can be written as
\begin{equation}
(E_{HF}^{[2n]}-E)|0\rangle_{\phi,\phi'}+\int_0^\pi d\phi' \mathcal{U}_{10}|1\rangle_{\phi,\phi'}=0,
\end{equation}
where $\mathcal{U}_{mn}={_{\phi,\phi'}\langle} m |\hat{\mathcal{U}}_L|n\rangle_{\phi,\phi'}$, and $E$ is the eigen energy. Considering the shorted path that connects $|0\rangle_{\phi,\phi'}$ and $|n\rangle_{\phi,\phi'}$ in Hilbert space as shown in Fig.(1) of the main text, the Schrodinger equation satisfied by $|m\rangle_{\phi,\phi'}$ can be approximated by
\begin{equation}
(E_m-E)|m\rangle_{\phi,\phi'}+\sum_{l=-1}^{1}\mathcal{U}_{m+l,m}|m+l\rangle_{\phi,\phi'}=0,
\end{equation}
where $E_m={_{\phi,\phi'}\langle} m |\hat{\mathcal{U}}|m\rangle_{\phi,\phi'}$ is the Hartree-Fock energy of $|m\rangle_{\phi,\phi'}$. Eliminating $|m\rangle_{\phi,\phi'}$ from the above equations, the effective Hamiltonian in Eq.(10) of the main text that takes the standard form of a high order perturbation theory is obtained.

We now turn to the diagonal term of $H_{eff}^{[2n]}$. The Hartree-Fock energy can be written as
\begin{equation}
E_{HF}^{[2n]}=\frac{1}{4}(6g_1+6g_2+2g_{12})n^2-\frac{1}{2}(g_1+g_2+g_{12})n, 
\end{equation}
For $\mathcal{E}'$, to the second order approximation for the diagonal term, we consider the lowest excited states with only one pair of atoms scattered out from the fragmented condensate and obtain 
 \begin{equation}
 \mathcal{E}'=\int_0^\pi d\phi' |\langle Ex |\hat{\mathcal{U}}_L| F^{[2n]}\rangle |^2/E_G,
 \end{equation}
 where $|Ex\rangle= \hat{L}^{\dagger n-1}_\phi \hat{L}^{\dagger n-1}_{\phi+\pi} \hat{L}^\dagger_{\phi'}\hat{L}^\dagger_{\phi'+\pi}|0\rangle/(n-1)!$ and $E_G =\langle Ex|\hat{\mathcal{U}}_L| Ex\rangle-E_{HF}^{[2n]}=(g_1+g_2+g_{12})(n-1)$. To be explicit, $\mathcal{E}'={\pi}\frac{n^2}{n-1}\frac{g_1^2+g_2^2}{g_1+g_2+g_{12}}$.\\

{\bf Solution of $H_{eff}^{[2n]}$}

For general cases of $g_1\neq0, g_2\neq0$, $\mathcal{V}_{\phi,\phi'}$ can be written as 
\begin{equation}
\mathcal{V}_{\phi,\phi'}=\frac{(-1)^{n-1}n^2}{(g_1+g_2+g_{12})^n}\sum_{\kappa=0}^{n}\mathcal{C}_n^{\kappa}g_1^{\kappa}g_2^{n-\kappa}e^{2i\kappa(\phi'-\phi)}.
\end{equation}
It is straightforward to verify that $|SF_{\nu}^{[2n]}\rangle$ are the eigen states with the eigen energies
\begin{equation}
E_{\nu}^{[2n]}=\Bigg\{\begin{array}{c}
\mathcal{E}_F^{[2n]}+\frac{(-1)^{n-1}n^2}{(g_1+g_2+g_{12})^{n-1}}({\pi \mathcal{C}_n^{\nu} g_1^{\nu}g_2^{n-\nu}-(g_1+g_2)^{n}})\\
 0\le \nu\le n\\\\
\mathcal{E}_F^{[2n]}-(-1)^{n-1}\frac{{(g_1+g_2)^{n}}n^2}{(g_1+g_2+g_{12})^{n-1}}\\ \nu<0, \nu> n\\
\end{array}\label{S4}
\end{equation}
where $\mathcal{C}_n^{\nu}=\frac{n!}{\nu !(n-\nu)!} $. For $N=2n=4l+2$, $(-1)^{n-1}>0$, all states with $\nu<0, \nu>n$ are degenerate and have lower energy than those with $0\le \nu\le n$. This means the ground states are infinitely degenerate. For $N=2n=4l$, $(-1)^{n-1}<0$, the ground state is unique, which is determined by the ratio of $g_1/g_2$ so that $\mathcal{C}_n^{\nu} g_1^{\nu}g_2^{n-\nu}$ is maximized. For example, for $g_1/g_2=0, 1, \infty$, the winding number for the ground state $\nu_0=0, n/2, n$ respectively. \\

{\bf Small $N$ systems}

The dependence of the ground state on the total particle number and interaction is summarized in Fig.(1). $\gamma_c(N)$ is the boundary between different phases, which is determined by comparing the energy of the single-particle condensate $|C^{[2n]}\rangle$ with other states. For instance, by solving $E_C^{[2n]}=E_\nu^{[2n]}$, one obtains the value of $\gamma_c(N)$ that separates $|C^{[2n]}\rangle$ and the super-fragmented state $|SF_\nu^{[2n]}\rangle$.

 \begin{figure}[tbp]
\begin{center}
\includegraphics[width=3.2in]{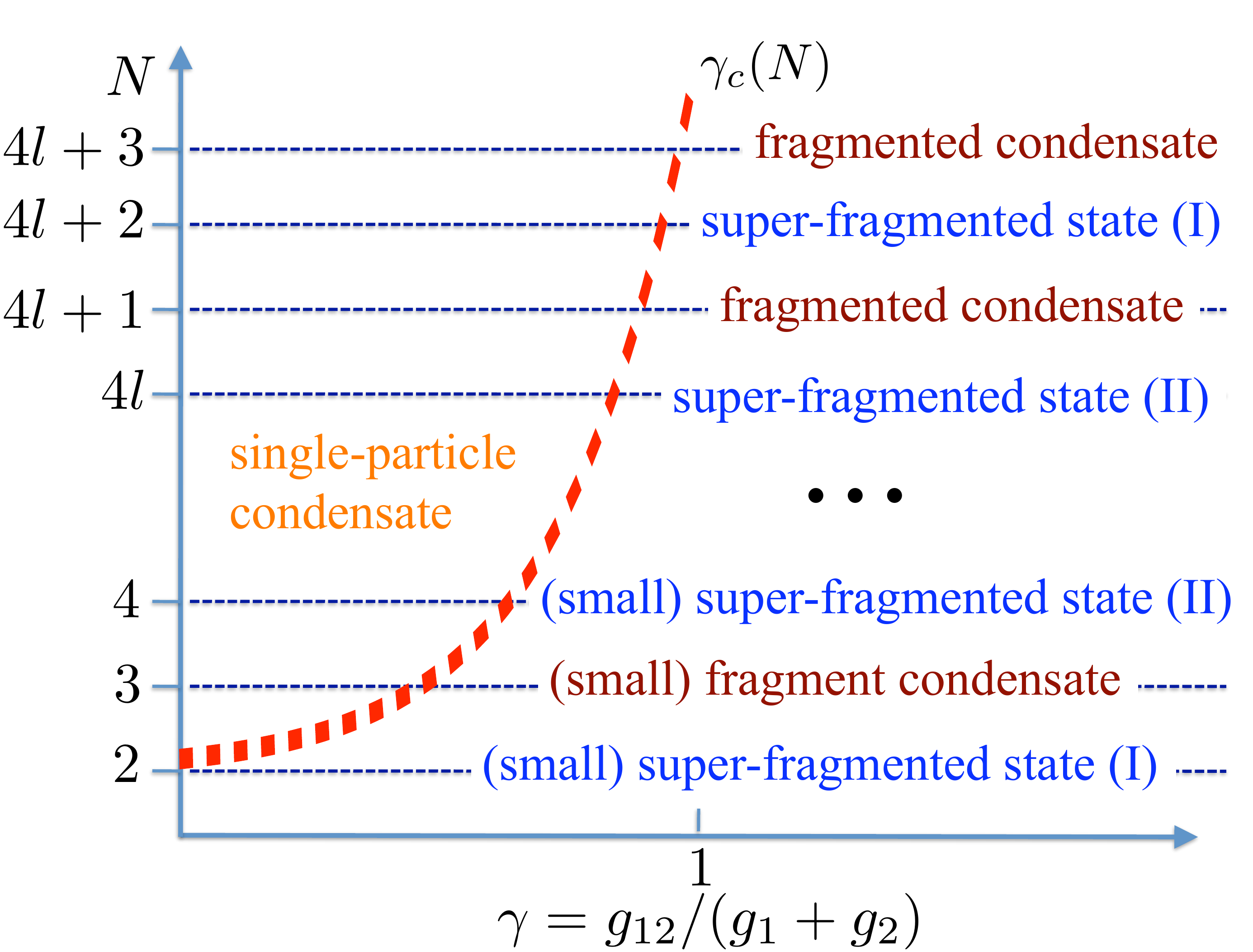}
\end{center}
\caption{Many-body ground states under Rashba coupling. single-particle condensate occupies the region with small $\gamma$. For large $\gamma$, the ground state is a fragmented condensate at odd particle number, $N=4l+1$ or $N=4l+3$, where $l$ is an integer. For even particle numbers, the ground state is a Super-fragmented state with infinite degeneracy for (I) with $N=4l+2$ and no degeneracy for (II) with $N=4l$. The dashed red curve represents the critical value $\gamma_c$ for the transition between single-particle condensate and other three states.  }
 \end{figure}

At small particle numbers, the derivation of $\gamma_c(N)$ from the mean field value $\gamma_c^0=1$ is significant. Particularly, as discussed in the main text, the small super-fragmented state is always the ground state for $N=2$, i.e., $\gamma_c(N)=0$.  With increasing $N$, the amplitude of the off-diagonal term decreases and the energy gained by the superposition of the fragmented states decreases. Therefore, $\gamma_c(N)$ increases from zero. It is  useful to study how $\gamma_c(N)$ approaches the mean field value.  Clearly, the contribution from $\mathcal{E}'$ to the ground state energy is much larger than that from off-diagonal term in Eq.(4) of the main text.   If one ignores $\mathcal{E}'$, $E_C^{[2n]}=\mathcal{E}_F^{[2n]}$ leads to the mean-field result $\gamma_c^0=1$. Taking into account $\mathcal{E}'$, the value of $\gamma_c$ is modified. In large $N$ limit, we find
\begin{equation}
\gamma_c(2n)\approx \gamma_c^0-\frac{\pi}{n}\frac{g_1^2+g_2^2}{(g_1+g_2)^2}.
\end{equation}
Similar scaling of $\gamma_c(N)-\gamma_c^0\sim 1/N$ holds for odd particle numbers. \\

{\bf Effect of anisotropy of SOC}

The kinetic energy offset in $\mathcal{L}'$ as shown in Eq. (8) of the main text leads to an additional term in the Hamiltonian for anisotropic SOC,
\begin{equation}
\hat{\mathcal{W}}=\int_0^\pi d\phi K(\phi)\hat{L}_{\phi}^\dagger\hat{L}_\phi.
\end{equation}
The matrix elements of $\hat{\mathcal{W}}$ between $|SF^{[2n]}_\nu\rangle$ can be calculated straightforwardly, 
\begin{equation}
\mathcal{W}_{\nu,\nu'}=\frac{N}{\pi}\int_0^\pi d\phi e^{2i(\nu-\nu')\phi} K(\phi)
\end{equation}
Since $K(\phi+\pi)=K(\phi)$, $\mathcal{W}_{\nu,\nu'}$ is finite only when $\nu-\nu'$ is an integer. For Rashba coupling, $K(\phi)$ is a constant and therefore $\mathcal{W}_{\nu,\nu'}=0$. With increasing $\delta K=K(\pi/2)-K(0)$ or equivalently the anisotropy of SOC  $\alpha$,  the amplitude of $\mathcal{W}_{\nu,\nu'}$ increases. 

The general expression of the ground state wave function can be written as 
\begin{equation}
|SF^{[2n]}\rangle_{\alpha}=\int_0^\pi d\phi C_\phi \hat{O}^\dagger_\phi|0\rangle.
\end{equation}
For isotropic SOC and $N=2n=4l$, $C_\phi=e^{2i\nu_0\phi}/\sqrt{\pi}$, and $|C_\phi|^2=1/\pi$ is a constant, where $\nu_0\le n$ as discussed before.  In the presence of a small anisotropy, $\hat{\mathcal{W}}$ is finite, one may take into account only the mixing between $|SF_{\nu_0}^{[2n]}\rangle$ and $|SF_{ \nu_0\pm 1}^{[2n]}\rangle$, which leads to a small variance of $|C_\phi|^2$. With increasing $\alpha$, more states with different $\nu$ are mixed into the ground state. As $\mathcal{W}_{\nu,\nu'}\approx N\delta K$, we see for large enough $N\delta K\gg E_{\nu>n}-E_{\nu_0}$,  all states $|SF_\nu^{[2n]}\rangle$ are mixed. An equal superposition of them leads to $C_{\phi}=\delta_{\phi, 0}/\sqrt{\pi}$ and the ground state becomes a fragmented condensate. 

For $N=2n=4l+2$, a subtle difference emergies due to the infinitely degenerate ground states under Rashba coupling. In the presence of any infinitesimal $\hat{\mathcal{W}}$, a superposition of  these degenerate state $|SF_{\nu>n}^{[2n]}\rangle$ may fulfill the goal of minimizing both the kinetic energy and interaction energy.  For instance, in the simplest case where $g_1=0$, the ground state becomes 
\begin{equation}
|SF^{[2n]}\rangle_{\alpha\rightarrow0}=\sum_{\nu\neq 0} |SF^{[2n]}_\nu\rangle=\int_0^{\pi} C_\phi\hat{O}^\dagger_\phi|0\rangle,
\end{equation}
where $C_\phi=({\pi} \delta_{\phi,0}-1)/\sqrt{\pi^2-\pi}$. With a finite $\alpha$, $|SF^{[2n]}_0\rangle$ will be mixed to the ground state. For large enough  $N\delta K\gg E_{\nu>n}-E_{\nu_0}$, similar to the case of $N=2n=4l$, all states $|SF_\nu^{[2n]}\rangle$ are mixed and a fragmented condensate becomes the ground state.\\

{\bf Fragmented condensate to a stripe condensate}

The fragmened condensate is a Fock state in the momentum space, with half particles occupying the state $\bf{k_0}$ and the other half occupying $-\bf{k_0}$. In homogenous systems, these two states can not be mixed due to momentum conservation. If an external potential exists and translational invariance is broken, an additional term exists in the Hamiltonian $H'=t'\hat{L}^\dagger_{\bf k_0}\hat{L}_{-\bf k_0}+c.c$, where $t'$ is proportional to the strength of the potential. Therefore, it becomes a double-well problem in momentum space. It is well known that the tunneling between the two ``wells" is enhanced with increasing particle number $N$ due to the bosonic enhancement effect, i.e., $H'\sim t' N$. In the thermodynamic limit, any infinitesimal $t'$ leads to the formation of a coherent state $\sim (\hat{L}^\dagger_{\bf k_0}+\hat{L}_{-\bf k_0}^\dagger)^N|0\rangle $, so called stripe condensate in the literature.

\end{document}